\documentclass[10pt]{iopart}
\expandafter\let\csname equation*\endcsname=\relax 
\expandafter\let\csname endequation*\endcsname=\relax 
\providecommand{\keywords}[1]{ #1}
\makeatletter
\def\ps@pprintTitle{%
 \let\@oddhead\@empty
 \let\@evenhead\@empty
 \def\@oddfoot{}%
 \let\@evenfoot\@oddfoot}
\makeatother
\renewcommand{\thetable}{\Roman{table}}
\usepackage{csquotes}
\usepackage{graphicx,color}
\usepackage{float}
\usepackage[caption=false]{subfig}
\usepackage{amsmath}
\usepackage{amssymb}
\usepackage{multirow}
\usepackage{dsfont}
\usepackage{adjustbox}
\usepackage{pdflscape}
\usepackage{wasysym}
\usepackage[percent]{overpic}
\usepackage[dvipsnames]{xcolor}
\usepackage[normalem]{ulem}
\begin{document}

\title{Impact of repetitive ELM transients on ITER divertor tungsten monoblock top surfaces}
\author{K. Paschalidis$^{1}$, S. Ratynskaia$^{1}$, P. Tolias$^{1}$, R. A. Pitts$^{2}$}
\address{$^1$Space and Plasma Physics - KTH Royal Institute of Technology, Teknikringen 31, 10044 Stockholm, Sweden\\
         $^2$ITER Organization, Route de Vinon-sur-Verdon, CS 90 046, 13067 St Paul Lez Durance Cedex, France}
\begin{abstract}
\noindent Owing to the high stored energy of ITER plasmas, the heat pulses due to uncontrolled Type I edge localized modes (ELMs) can be sufficient to melt the top surface of several poloidal rows of tungsten monoblocks in the divertor strike point regions. Coupled with the melt motion associated with tungsten in the strong tokamak magnetic fields, the resulting surface damage after even a comparatively small number of such repetitive transients may have a significant impact on long-term stationary power handling capability. The permissible numbers set important boundaries on operation and on the performance required from the plasma control system. Modelling is carried out with the recently updated MEMENTO melt dynamics code, which is tailored to tackle melt motion problems characterized by a vast spatio-temporal scale separation. The crucial role of coupling between surface deformation and shallow angle heat loading in aggravating melt damage is highlighted. As a consequence, the allowable operational space in terms of ELM-induced transient heat loads is history-dependent and once deformation has occurred, weaker heat loads, incapable of melting a pristine surface, can further extend the damage.
\end{abstract}
\maketitle
\ioptwocol

\section{Introduction}\label{sec:introduction}

The sufficiently long lifetime of plasma-facing components (PFCs) constitutes a critical issue for the development of future fusion reactors. The ITER tokamak will be equipped with a full-tungsten (W) divertor using actively cooled monoblock (MB) technology capable of handling high stationary thermal loads\,\cite{Pitts2017}. Rapid and energetic transient events such as edge-localized modes (ELMs) can severely compromise divertor MB longevity as a result of repetitive, shallow top surface melting, associated recrystallization deeper into the material and possible eventual macro-cracking of the MB. Once MB top surface melting occurs, melt motion\,\cite{Ratynskaia2020} alters their carefully engineered shape, making them more susceptible to damage under subsequent thermal loading, either transient or stationary. Macroscopic melt motion may also lead to the bridging of gaps between MBs or castellations introduced into PFCs to suppress the flow of eddy currents. Thus, temperatures above a certain threshold must be avoided, depending on the time over which the temperature is sustained (recrystallization is a function of time and temperature)\,\cite{Tsuchida2018}. Finally, concerning the integrity of the cooling system, the temperatures at the coolant interface must be kept below a given threshold\,\cite{Escourbiac2019}. Exceeding the threshold for a long enough time will lead to wall critical heat flux (CHF) events, where the cooling system fails to remove heat from the MB due to vapor formation.

Previous computational efforts in the area of MB loading have been mostly devoted to studies of steady-state loading tolerances, investigations of the cooling system limits and modelling of recrystalization as well as recrystalization-driven cracking  \cite{Escourbiac2019,Hirai2016,Hirai2018,Panayotis2017,Gunn2017,Van_den_Kerkhof2021}. In scenarios where melting takes place, the melt motion can introduce significant changes to the MB thermal response by altering the topology of the top surface and hence the distance to the coolant. Furthermore, in the case of top surface loading, the magnetic field, and thus the heat flux, is incident with a shallow angle. In such cases, there is a pronounced coupling between deformation and thermal loading; as the top surface evolves due to macroscopic melt motion so does the local magnetic field incidence angle. Such a coupling can lead to striations in the deformation profile, as observed in experiments with a 18$^\circ$ slopped W sample exposed in the ELMing H-mode plasmas of ASDEX Upgrade\,\cite{Krieger2018} and as reproduced in MEMOS-U simulations of this experiment\,\cite{Thoren2018}.

Modelling the macroscopic melt motion induced on an actively cooled ITER MB by ELM heat loads poses a significant computational challenge. The complexities in the MB geometry, the evolution of the free-surface, the different temporal scales imposed by the long ($\sim 1$\,s) inter-ELM periods interrupted by short ($\sim 1$\,ms) bursts of intense intra-ELM fluxes as well as the separation of spatial scales (tens of $\mu$m melt depths versus $\sim 1$ cm-long MB) require highly adaptive meshing and time steps. In addition, for accurate estimates of the angle of incidence of the heat flux on the deformed surface, the effect of the surface tension must be considered due to its role in the relaxation of the deformation curvature.

Such simulations were not feasible until very recently. Here, capitalizing on numerical updates to the MEMENTO melt dynamics code\,\cite{Paschalidis2023,Paschalidis2024}, we address the problem of ITER MB top surface damage under repetitive ELM transient heat loads. MEMENTO is a computational tool developed specifically to predict macroscopic melt motion in harsh fusion reactor environments\,\cite{Paschalidis2024} whose physics model has been validated against several dedicated melt experiments on a range of devices, within EUROfusion and ITPA DivSOL Topical Group multi-machine tasks \cite{Ratynskaia2020,Thoren2018,Thoren2017,Thoren2018_1,Thoren2021,Ratynskaia2021,Ratynskaia2022,Ratynskaia2024}. The aim of this work is to scope the thresholds and allowed number of uncontrolled ELM transients before damage occurs which may affect the long-term performance of the ITER divertor MBs. In particular, we explore the ELM parameter space to identify combinations where CHF events, possible recrystallization, non-negligible melting and surface deformation take place.

\section{Statement of the problem}\label{sec:problem}

The melt damage extent will be assessed in terms of the excavation volume and the onset of the liquid-solid phase transition. An MB will be considered to have endured significant recrystallization damage when the temperature $2\,$mm below the center of its top surface reaches $2000\,$K. The $2000\,$K temperature lies near the upper threshold for the recrystallization temperature (which varies roughly within $1450-2050\,$K depending on the manufacturing\,\cite{Smid1998,Davis1998}). \textcolor{black}{Note that the grain structure of pure W requires $\sim 1$\,h to change at such temperatures\,\cite{Tsuchida2018}}. The $2\,$mm depth has been associated with the emergence of macro-cracks\,\cite{Li2015} and is used in Refs.\cite{Li2017, Pitts2019} as the basis for estimating the allowed stationary power loading of ITER MB's. \textcolor{black}{Finally, the chosen criterion for a CHF occurrence is that the temperature at the coolant interface reaches $573$\,K. It should be noted that there is some uncertainty concerning this limit and that the CHF conditions might be met at lower temperatures. Detailed investigations of the CHF limits can be found in Refs.\cite{Raffray1999, Boscary1999}.}

The MEMENTO code solves heat transfer including phase change coupled with the incompressible Navier-Stokes equations in the shallow water approximation, also accounting for current propagation, on a domain with a time-evolving deforming metal-plasma interface\,\cite{Paschalidis2023, Paschalidis2024, Ratynskaia2022, Ratynskaia2024}. The plasma heat flux constitutes an external input for MEMENTO and is specified in Sec.\ref{ssec:hf}. In this work, all relevant momentum sources for the melt dynamics have been included; the Lorentz force induced by the replacement current triggered by thermionic emission, the thermo-capillary force due to presence of surface temperature gradients and the surface tension (gravity is also included, but it is negligible). The inclusion of surface tension is necessary for better estimates of the local surface normal.

\subsection{Geometry}\label{ssec:geometry}

The details of the MB geometry can be found in Ref.\cite{Hirai2018}. It comprises W, a Cu interlayer and a CuCrZr cooling pipe. The MB dimensions are $28\times28\times12$\,mm$^3$ and its top surface is toroidally bevelled at a $\sim1^\circ$ angle, aiming to shadow any leading edges due to radial misalignments between toroidally neighbouring MBs and ensure that only the top surface is wetted by the plasma. \textcolor{black}{In this work, the full top surface of the MB is wetted, corresponding to the worst-case scenario.}

\subsection{Heat flux}\label{ssec:hf}

The adopted ELM heat flux incident on a MB at the outer divertor is based on Refs.\cite{Loarte2003, Loarte2007, Loarte2014, Fundamenski2006, Eich2017} which are combined in the same manner as in Ref.\cite{Van_den_Kerkhof2021}. For the sake of completeness, the procedure is briefly repeated here. The peak parallel energy fluence (in MJ/m$^2$) is obtained from the Eich scaling\,\cite{Eich2017}
\begin{multline}
    \label{eqn:eich}
\epsilon_{\|,\mathrm{ELM}}^{\mathrm{P}}=0.28\pm0.14\times{n}_{\text{e,ped}}^{0.75\pm0.15}T_{\text{e,ped }}^{0.98\pm0.1}
    \\
    \Delta{E}_{\mathrm{ELM}}^{0.52\pm0.16}R_{\text{geo}}^{1\pm0.4},
\end{multline}
where \cite{Loarte2014}
\begin{equation}\label{eqn:SOL2ELM}
    \begin{aligned}
        &\Delta E_{\mathrm{ELM}} = \Delta W_{\mathrm{ELM}}/W_{\mathrm{plasma}}\  [\%], 
        \\
        &\Delta W_{\mathrm{ELM}} = 0.3 P_{\mathrm{SOL}}/f_{\mathrm{ELM}}.
    \end{aligned}
\end{equation}
with $n_{\text{e,ped}}$ the pedestal top density in $[10^{20}\,\mathrm{m}^{-3}]$, $T_{\text{e,ped}}$ the pedestal top electron temperature in $\mathrm{keV}$, $\Delta E_{\mathrm{ELM}}$ the ELM energy loss relative to the plasma stored energy in percentage, $R_{\mathrm{geo}}$ the major radius in m, $P_{\mathrm{SOL}}$ the scrape-off layer (SOL) power and $f_\mathrm{ELM}$ the ELM frequency in Hz. Note that the factor 0.3 in Eq.\,\eqref{eqn:SOL2ELM} is taken as an average of the range (0.2 - 0.4) of the fraction of $P_\mathrm{SOL}$ found to be exhausted by uncontrolled Type I ELMs in current devices.

The time dependence is described by 
\begin{equation}
    \frac{q_{\perp, \mathrm{ELM}}^{\mathrm{P}}(t)}{q_{0, \mathrm{ELM}}}=\left[1+\left(\frac{\tau}{t}\right)^{2}\right]\left(\frac{\tau}{t}\right)^{2} \exp{\left[-\left(\frac{\tau}{t}\right)^{2}\right]},
\end{equation}
where $\tau=0.8\tau_{\mathrm{IR}}$ with $\tau_{\mathrm{IR}} = 250\,\mathrm{\mu s}$ the time constant of the rise phase expected for Type I ELMs in ITER burning plasmas at 15 MA\,\cite{Loarte2003, Loarte2007, Fundamenski2006}. The scaling coefficient $q_{0, \mathrm{ELM}}$ is found via 
\begin{equation}
    \epsilon_{\perp, \mathrm{ELM}}^{\mathrm{P}}=\int_{\tau_{\mathrm{ELM}}} q_{\perp, \mathrm{ELM}}^{\mathrm{P}}(t) d t,
\end{equation}
with $\tau_{\mathrm{ELM}}=f^{-1}_{\mathrm{ELM}}$ the ELM period and with $\epsilon_{\perp,\mathrm{ELM}}^{\mathrm{P}}=\epsilon_{\|,\mathrm{ELM}}^{\mathrm{P}}\sin{(a_0)}$ for the optical connection between the peak perpendicular and parallel energy fluences, where $a_0=4.2^\circ$ is the heat flux angle of incidence on the top surface MB at the outer target\,\cite{Pitts2019}. Finally, the spatial variation of the heat flux is neglected due to the weak gradients over the MB poloidal length (12\,mm) and due to the assumption of toroidal symmetry.

\subsection{Plasma conditions}\label{ssec:plasma_conditions}

The ITER parameters for Eq.\eqref{eqn:eich} are assumed to be $n_{\mathrm{{e,ped}}} = 8.9\times10^{19}\,\mathrm{m}^{-3}$, $T_{\mathrm{{e,ped}}}=4.7\,\mathrm{keV}$, $W_\mathrm{plasma}=350\,$MJ, $P_\mathrm{SOL}=100$\,MW and $R_\mathrm{geo}=6.2\,$m. Furthermore, MEMENTO requires an estimate for the value of the plasma density at the target for the analytical modelling of the space-charge limited thermionic emission. Following Ref.\cite{Ratynskaia2021}, the plasma density can be expressed through the heat flux by assuming $T_{\mathrm{e}}=T_{\mathrm{i}}=0.5T_{\mathrm{{e,ped}}}$, a sheath heat transmission coefficient of $\gamma=7$ and that ions enter the emissive sheath with the Bohm speed. The maximum parallel heat flux for $f_\mathrm{ELM}=10\,$Hz and at the upper limit of Eich scaling uncertainties is employed to obtain $n_{\mathrm{e}} \sim 1.1\times10^{20}\,\mathrm{m^{-3}}$. As discussed in section \ref{ssec:implem_loading}, the dependence of thermionic emission on the density is rather weak, thus allowing the use of a single constant density value for all simulations without introducing errors.

\section{Implementation in MEMENTO}\label{sec:implementation}

\subsection{Geometry and active cooling}\label{ssec:implem_geometry}

\begin{figure}
    \centering
    \includegraphics[width=0.9\linewidth]{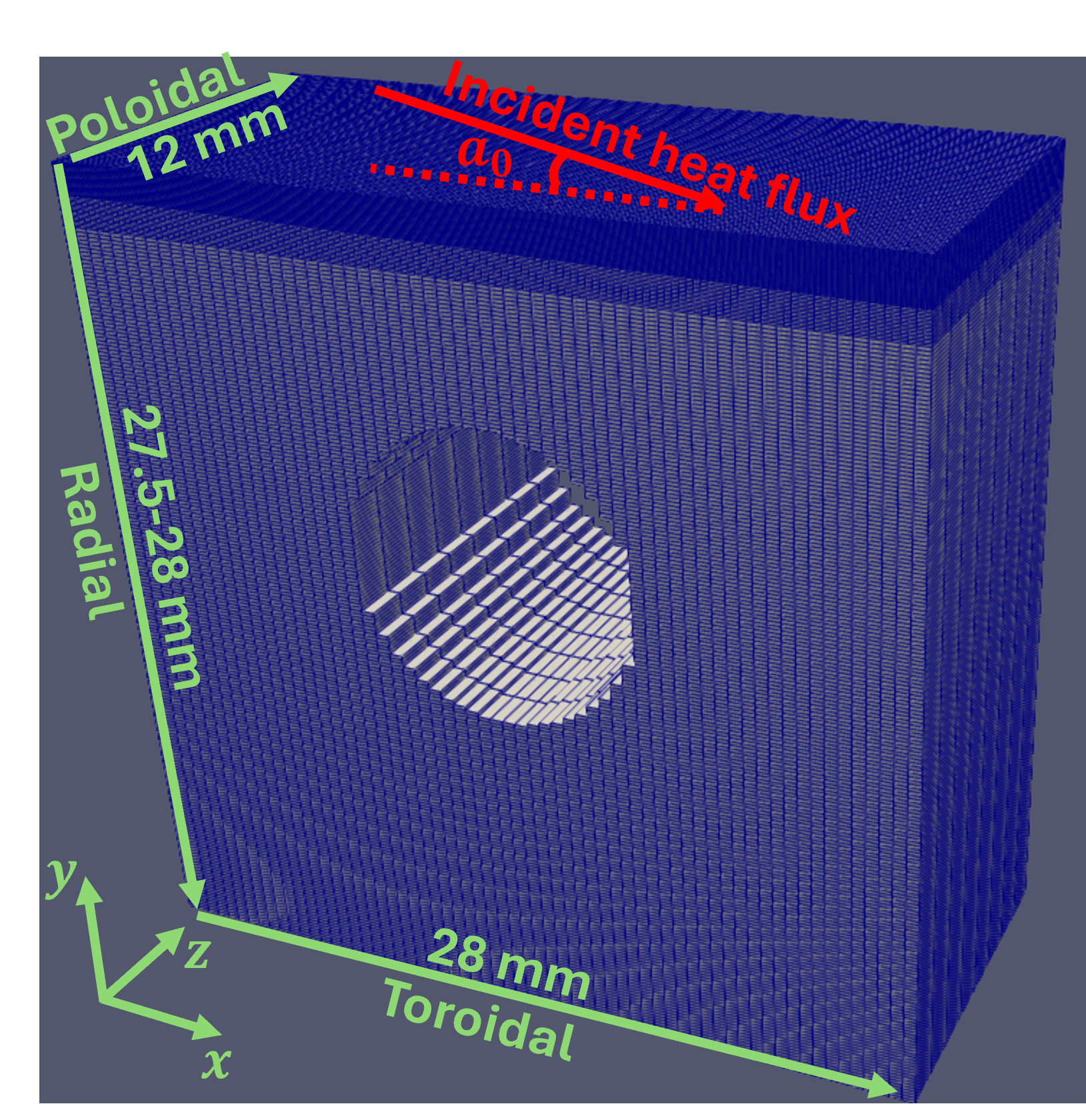}
    \caption{The geometry, coordinate system and MEMENTO grid. Note that the radial length is $27.5\,$mm at $x=0$\,mm and $28$\,mm at $x=28$\,mm due to the toroidal bevel. The angle between the incident heat flux and the surface tangent to the pristine surface is also sketched.}
    \label{fig:3D_geom}
\end{figure}

The geometry is mapped onto the rectilinear grid of the MEMENTO code, as shown in Fig.\ref{fig:3D_geom}. Simplifications are followed in the cooling pipe implementation, that were first introduced in Ref.\cite{Gunn2017}. To be more specific, the cooling pipe radius is reduced to $5\,$mm, the $2.5\,$mm of CuCrZr and Cu are substituted by $1\,$mm of W and the cooling pipe is shifted so that there are $7\,$mm of W between the middle of the top surface and the coolant. It has been demonstrated that such an implementation provides reasonable accuracy; in Ref.\cite{Paschalidis2023} comparison between MEMENTO and ANSYS simulations of the MBs revealed minimal discrepancies, while in Ref.\cite{Ratynskaia2022}  MEMENTO successfully reproduced the results of a controlled melting experiment carried out at WEST on an ITER-like MB.

The coolant is assumed to flow at $100^\circ\,$C and the heat transfer coefficient originally presented in Ref.\cite{Gunn2017} is employed. For coolant interface temperatures that exceed the $\sim570$\,K upper validity threshold\,\cite{Gunn2017}, the coefficient can be safely assumed to be constant, since no simulation exceeded this limit by more than $\sim30\,$K.

\subsection{Loading scenarios}\label{ssec:implem_loading}

Different loading scenarios were constructed by varying $f_\mathrm{ELM}$ and the uncertainty-prone parameters (pre-factor and exponents) of the Eich scaling, see Eq.\eqref{eqn:eich}. ELM frequencies in the range of $1\,$Hz to $50\,$Hz were explored. For any $f_{\mathrm{ELM}}$, the best case, corresponding to the lower uncertainty limits, does not result in any melting and is thus not considered here. Therefore, for any $f_{\mathrm{ELM}}$, we will explore: (i) the nominal case, (ii) the worst case corresponding to the upper uncertainty limits.

\begin{figure}
    \centering
    \includegraphics[width=0.9\linewidth]{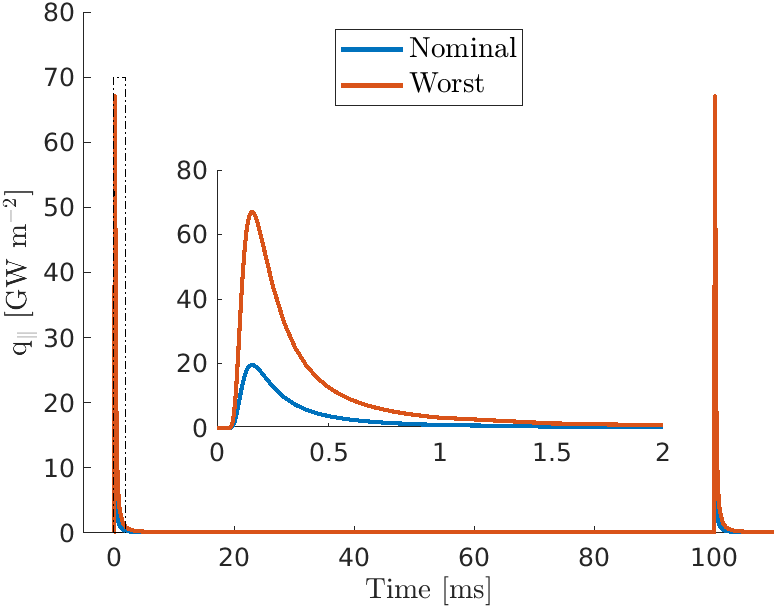}
    \caption{The parallel heat flux for the nominal and worst cases; $f_\mathrm{ELM} = 10$\,Hz.}
    \label{fig:elm-flux}
\end{figure}

At a representative frequency of 10\,Hz, Fig.\ref{fig:elm-flux} features the time trace of the parallel heat flux for these two cases. It illustrates the temporal scale separation arising from the rapid increase of the intra-ELM flux and the long waiting period between ELMs. Note that the worst case is equivalent to using nominal values but increasing $\Delta E_\mathrm{ELM}$ from $\sim$1\% to $\sim$10\%

In addition, an inter-ELM heat flux must be specified. Different values were employed, a moderate  $q_\parallel^\mathrm{inter}=6/\sin{a_0}\,\mathrm{[MW m}^{-2}]$ and one slightly above the defined ITER tolerances $q_\parallel^\mathrm{inter}=12/\sin{a_0}\,\mathrm{ [MW m}^{-2}]$. 

This procedure defines the field line parallel heat flux which is then projected onto the MB top surface to yield the heat flux loading. As prescribed above, the heat flux incident angle on the pristine bevelled surface is 4.2$^\circ$. However, when melt motion occurs, the local surface normal at the MB top surface will continuously evolve due to deformation and hence the projected heat flux constantly changes in the course of the MEMENTO simulation.

To model the variation of the projected area, the unit normal vector to the free surface is introduced. The heat flux is split into a fraction $F_{\mathrm{o}}$ that follows the optical approximation and the remaining 1-$F_{\mathrm{o}}$ which is independent of deformation, as prescribed in Ref.\cite{Thoren2018}. The projected heat flux is given by 
\begin{equation}
  \label{eqn:par2perp}
  q_\perp = \left\{
    \begin{array}{ll}
      F_{\mathrm{o}} q_\parallel \sin{a} + (1-F_{\mathrm{o}}) q_\parallel \sin{a_0}, &a\leq \pi \\
      (1-F_{\mathrm{o}}) q_\parallel \sin{a_0}, &a>\pi, 
    \end{array}
  \right.
\end{equation}
with $a_0=4.2^{\circ}$ being the angle between the heat flux vector and the pristine surface tangent plane. Similarly, $a$ is the angle between the heat flux vector and the tangent to the deformed surface. Note that $a$ evolves in space and time, while $a_0$ is constant. The weight $F_{\mathrm{o}}$ is set to $2/7$, approximating the fraction of the heat deposited by electrons. The ion Larmor radius is $\sim1$\,mm for the assumed pedestal top parameters and the local B-field strength ($\sim7$\,T). The expected characteristic deformation length scale is also $\sim1$\,mm and hence the ion contribution cannot be modelled by the optical approximation. 

The heat flux projection introduces complexities in the calculation of the escaping current density due to thermionic emission. In MEMENTO, the escaping current is computed as $J_\mathrm{esc} = \min(J^\mathrm{nom}_\mathrm{th}, J^\mathrm{lim}_\mathrm{th})$, where $J^\mathrm{nom}_\mathrm{th}(T_{\mathrm{s}})$ is the Richardson-Dushmann current that is independent of the plasma conditions and where $J^\mathrm{lim}_\mathrm{th}$ is the maximum escaping current within the space-charge limited regime
that is practically independent of the surface temperature\,\cite{Thoren2021}. Systematic PIC simulations of magnetized emissive sheaths have revealed that the limited value of the escaping current density obeys the scaling $J^\mathrm{lim}_\mathrm{th}\propto q^{1/3}n^{2/3}\sin^2{a}$\,\cite{Komm2017,Komm2017_1,Komm2020,Tolias2020,Tolias2023}. Apart from the assumption of homogeneous boundary conditions, these PIC simulations considered perfectly planar W surfaces. In practice, the escaping current density and even the plasma conditions in front of the plasma shadowed regions are uncertain. The effect of this is further explored in Sec.\ref{sssec:thermionic}. It is also pointed out that secondary electron emission and electron backscattering are not taken into account, even though the associated yields should be very high at the relevant incident energies and angles\,\cite{Tolias2023,Tolias2014_1,Tolias2014}. Essentially, it is assumed that, given the grazing angles and strong magnetic fields, prompt re-deposition of the emitted electrons will be very effective\,\cite{Komm2020}. Nevertheless, it is worth pointing out that strong electron backscattering would imply that a respectable amount of the incident ELM energy returns to the edge plasma\,\cite{Tolias2014}.

\subsection{Replacement current}\label{ssec:electro}

The bending of the replacement current density, which is triggered by the escaping thermionic current and is responsible for the Lorentz force density, depends on the sample and loading geometries\,\cite{Thoren2018_1}. For shallow angle loading, the replacement current density is nearly perpendicular to the external magnetic field, while for leading edge loading, only a minor current component is not parallel to the magnetic field\,\cite{Thoren2018_1}. Self-consistent evaluation of current propagation requires the numerical solution of an elliptic partial differential equation which significantly affects the computational time\,\cite{Paschalidis2023, Paschalidis2024}. Therefore, it is important to assess whether the electrostatic module of MEMENTO can be circumvented in the scenarios considered, without affecting the overall accuracy of the simulations.  

In the present ITER simulations, the inclination angle of the magnetic field is initially grazing to the surface but continuously evolves due to deformation. Simulations revealed that the y-component of the local normal (see the coordinate system in Fig.\ref{fig:3D_geom}) is above $ \sim 0.8$ for almost all spatial points and times. For worst-case predictions, the replacement current density in each melt column can be considered to be equal to the escaping current density with a direction perpendicular to the magnetic field, introducing errors in the Lorentz force density that do not exceed $\sim20$\%. Thus, we conclude that a simplified cost-effective treatment of the replacement current can be pursued, since the associated uncertainty is smaller than the ELM heat flux uncertainty.

\section{Results}\label{sec:results}

For the present ITER scenarios, the full 3D simulations are computationally expensive due to the separation of temporal scales, the thin melt pools and the presence of the cooling pipe, which necessitates the use of a relatively fine grid even far away from the free surface. In the absence of melt motion, the problem essentially becomes 2D, due to the absence of gradients along the poloidal direction, which implies a drastic reduction in the computational cost. Thus, we first scan the entire parameter space with 2D MEMENTO heat transfer simulations to identify possible melting events and we then investigate the chosen loading scenarios with melt motion enabled in full 3D.

\subsection{2D scoping simulations without melt motion}\label{ssec:2d_runs}

In the 2D runs reported in this subsection, the inter-ELM flux was set to $6\,\mathrm{MW m}^{-2}$.

\subsubsection{Nominal case.}

Two runs were performed with nominal ELM heat flux coefficients, one with $f_\mathrm{ELM}=1$\,Hz and one with $f_\mathrm{ELM}=10$\,Hz. 

The $1\,$Hz simulation revealed superficial melting ($\sim8\,\mu$m deep) that occurs only after $3\,$s of exposure. In this case, the surface deformation induced by melt motion can be evaluated by analytical estimates. The shallow water averaged momentum equation can be employed for upper-limit estimates of the displacement by neglecting all spatial gradients.  Consideration of the Lorentz force and the omnipresent viscous damping yields the terminal velocity $u\approx0.1\,$m/s. When heat advection is insignificant, as is the case here due to the absence of spatial gradients, this terminal velocity is promptly reached over $\tau = (3 \mu/h^2)^{-1}$ timescales, where $h$ is the melt depth and $\mu$ the dynamic viscosity. From the conservation of mass in the shallow water approximation (the column height equation, see e.g. Eq.(1) in Ref.\cite{Paschalidis2023}), the characteristic length for the free surface position change $\delta b_2 = {u h \tau_\mathrm{lt}}/{L}$ is obtained, where $\tau_\mathrm{lt}$ is the pool life-time and $L$ the characteristic tangential length. With $\tau_\mathrm{lt}\sim 500\ \mu$s and $L$ equal to the MB poloidal length (12 mm), the deformation is barely $\delta b_2 = 1$ nm.  Hence, this case is not deemed as interesting enough to be self-consistently simulated with melt motion.

The $10\,$Hz simulation revealed that no melting is realized. The study in Ref.\cite{Van_den_Kerkhof2021} confirmed that further increasing $f_\mathrm{ELM}$ would not lead to any melting, since the maximum MB  temperature is reduced for higher ELM frequencies. Such simulations would be useful only to investigate CHF events, since the average ELM heat flux over $\tau_\mathrm{ELM}$ scales as $\propto f_\mathrm{ELM}^{0.48\pm 0.16}$ with the ELM frequency, obtained by substituting Eq.\eqref{eqn:SOL2ELM} into Eq.\eqref{eqn:eich}\,\cite{Van_den_Kerkhof2021}. However, as will be demonstrated below, even in the worst case, the CHF is only marginally reached at $50\,$Hz.

\subsubsection{Worst case.}

Simulations with ELM frequencies of 5,\,10,\,15,\,20 and 50\,Hz were run. No simulation with 1\,Hz was performed, since it was shown in Ref.\cite{Van_den_Kerkhof2021} that in this case the surface temperature reaches $\sim9000$\,K and vapour shielding would need to be taken into account to obtain realistic results\,\cite{Skovorodin2016}.

\begin{figure}
    \centering
    \subfloat{%
      \begin{overpic}[width =1.0\linewidth]{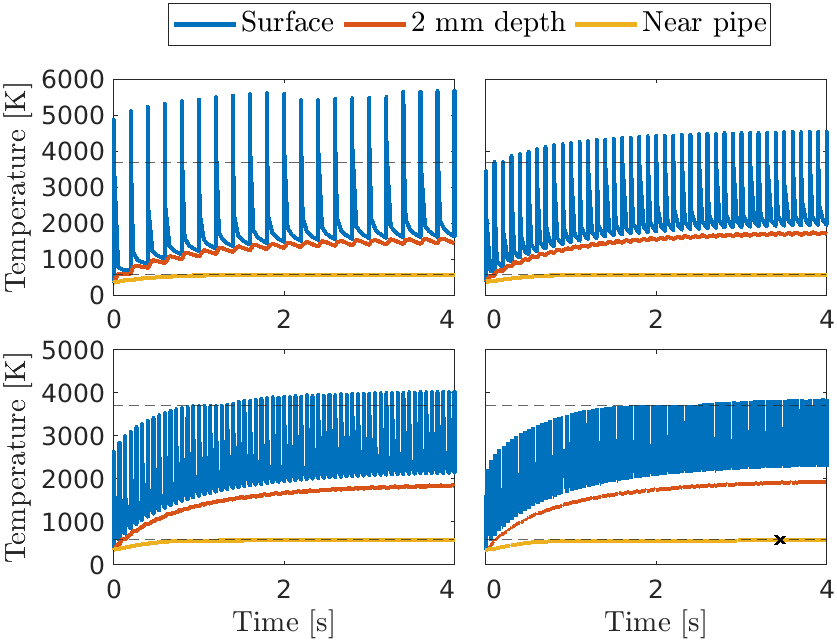}
        \put(42,44.5){\textcolor{black}{\textbf{5 Hz}}}
        \put(63,62){\textcolor{black}{\textbf{10 Hz}}}
        \put(15,30){\textcolor{black}{\textbf{15 Hz}}}
        \put(63,30){\textcolor{black}{\textbf{20 Hz}}}
      \end{overpic}          
    }
    \caption{Results of 2D runs (without melt motion) for the worse case loading scenario and for $f_\mathrm{ELM}=5,\,10,\,15,\,20$\,Hz. Blue solid lines for the maximum surface temperature, red solid lines for the temperature at $2\,$mm below the free surface at the center of the pipe and yellow solid lines for the temperature at the top of the cooling pipe. The melting and CHF temperatures are marked with dashed lines. The cross in the 20 Hz plot marks the instant when the CHF temperature is reached.}
    \label{fig:2D}
\end{figure}
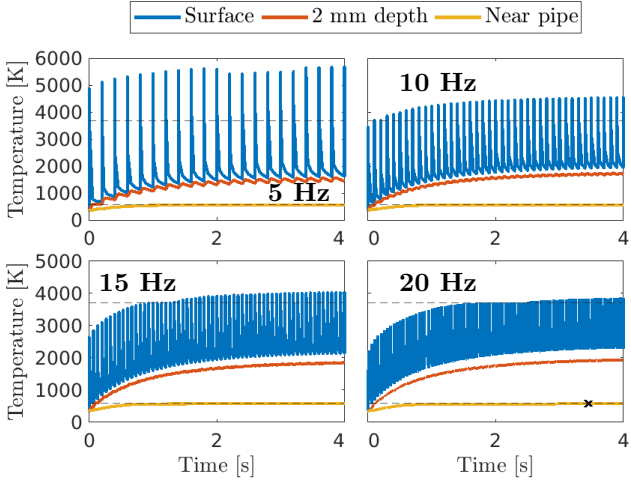

The 50\,Hz simulation showed that the CHF limit is reached $\sim2\,$s prior to the onset of melting. The temperature near the coolant plateaus at $\sim600$\,K, above the CHF limit. A summary of the findings for the other frequencies is shown in Fig. \ref{fig:2D}. At $f_\mathrm{ELM} = $ 15 and $20\,$Hz, the temperature $2\,$mm below the free surface reaches $\sim2000$\,K. Moreover, in both these cases the temperature at the interface to the coolant is within $2\,$ K of the CHF value. For 5 and $10\,$Hz neither the CHF limit nor $2000\,$K at $2\,$mm below the MB top surface are reached. The melt depth is $\sim65\,\mu$m and $\sim35\,\mu$m for $5\,$Hz and $10\,$Hz respectively and the pool lifetimes are $\sim\,1$ms. 

\begin{figure}
    \centering
    \subfloat{%
      \begin{overpic}[width =1.0\linewidth]{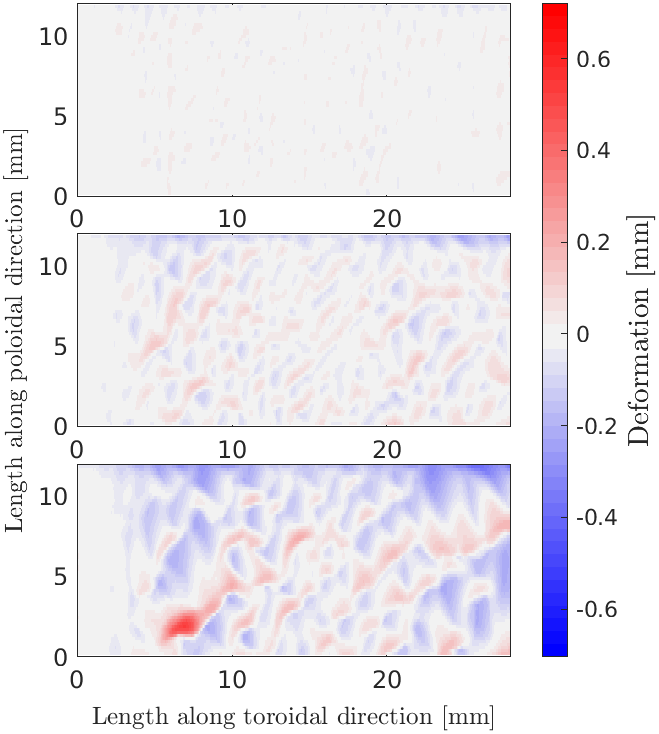}
        \put(13,75){\textcolor{black}{\textbf{\Large{2 s}}}}
        \put(13,43){\textcolor{black}{\textbf{\Large{3 s}}}}
        \put(13,11){\textcolor{black}{\textbf{\Large{4 s}}}}
      \end{overpic}          
    }
    \caption{Results of 3D runs (with melt motion as well as with coupled heat flux and deformation) for the worse case loading scenario, $f_\mathrm{ELM}=10\,$Hz and $q_\parallel^{\mathrm{inter}} = 6/\sin(a_0)$\,[MWm$^{-2}$]. The surface deformation profile is plotted after $2,\,3,\,4\,$s.}
    \label{fig:deformation evolution}
\end{figure}

Overall, for 5 and $10\,$Hz, the melt damage can amount to significant surface deformation. Due to the high computational cost of these runs, we chose only 10 Hz for 3D simulations with melt motion.

\subsection{3D simulations with melt motion}\label{ssec:3d_runs}

In this section, we start with the worst case for $f_\mathrm{ELM}=10\,$Hz and $q_\parallel^{\mathrm{inter}} = 6/\sin(a_0)$\,[MWm$^{-2}$]. This is the reference case which will be used to test the sensitivity of results to the differences in loading (with and without coupling between the thermal loading and deformation, see Secs.\ref{sssec:coupling},\,\ref{sssec:constant_loading}), to the 
uncertainties in the escaping thermionic current in the shadowed regions (see Sec.\ref{sssec:thermionic}) and to the variations of the inter-ELM values (see Sec.\ref{sssec:interELM}).

\subsubsection{Loading with coupled heat flux and deformation.}\label{sssec:coupling}

It should be emphasized that once melting occurs, the whole top surface is molten. In the simulations, the  deformation is initially seeded mainly by irregularities of the free surface due to the mapping of the bevelled surface onto the rectilinear grid of MEMENTO. In reality, even in the absence of surface roughness, an irregular deformation profile would emerge due to the varying melt thickness owing to the different distances of the solid-liquid interface from the cooling pipe and owing to the surface tension mainly at the MB edges. Once the top surface becomes non-planar, irregularities further drive deformation, creating a positive feedback loop. It is important to stress that only the early stages of deformation are affected by the rectilinear grid, since deformations quickly emerge that are characterized by length scales much larger than those arising from the discretization.

For the reference case, the evolution of the surface deformation is presented in Fig.\ref{fig:deformation evolution} starting from the 2\,s instant when noticeable deformation is first reached. Cavities (craters), resulting from mass displacement, are illustrated in blue, while material build-up is shown in red. The volume of the craters grows non-linearly in time; being 2.3\,mm$^3$, 7.4\,mm$^3$ and 18\,mm$^3$ at 2, 3 and 4\,s, respectively. Material accumulation evolves faster over time not only because the MB heats up (as observed in Fig.\ref{fig:2D}, without motion, it is nearly at steady-state already after 3\,s), but predominately because the projected heat flux is sensitive to the local surface normal, which increases with deformation. 

\begin{figure}
    \centering
    \includegraphics[width=0.88\linewidth]{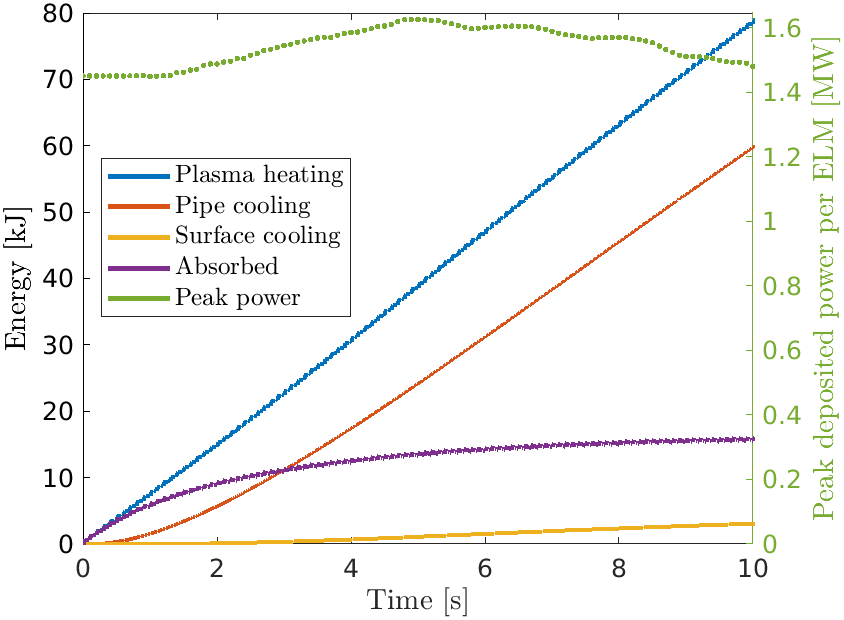}
    \caption{Results of 3D runs (with melt motion as well as with coupled heat flux and deformation) for the worse case loading scenario, $f_\mathrm{ELM}=10\,$Hz and $q_\parallel^{\mathrm{inter}} = 6/\sin(a_0)$\,[MWm$^{-2}$]. The energy balance (solid lines, left axis) and the peak power deposited in each ELM (dotted line, right axis) \textcolor{black}{evaluated as $\max_{\tau_\mathrm{ELM}}{\left\{\iint q_\perp(x,z,t) dx dz\right\}}$.}}
    \label{fig:energy_balance}
\end{figure}

This is also reflected in the temporal evolution of the deposited energies shown in Fig.\ref{fig:energy_balance}. The peak power per ELM, evaluated as $\max_{\tau_\mathrm{ELM}}{\left\{\iint q_\perp(x,z,t) dx dz\right\}}$, is changing in time only due to deformation and its higher values translate to higher melt damage per ELM. Note that the peak power per ELM starts to decline after $\sim 5$\,s but the deformation build-up does not slow down. This highlights how integrated quantities are indicative but not sufficient to describe the dynamics. The peak power might be declining, mainly due to the formation of shadowed regions, but {\it locally} the heat flux, which is the quantity affecting melt motion, does not.

\begin{figure}
    \centering
    \includegraphics[width=1.0\linewidth]{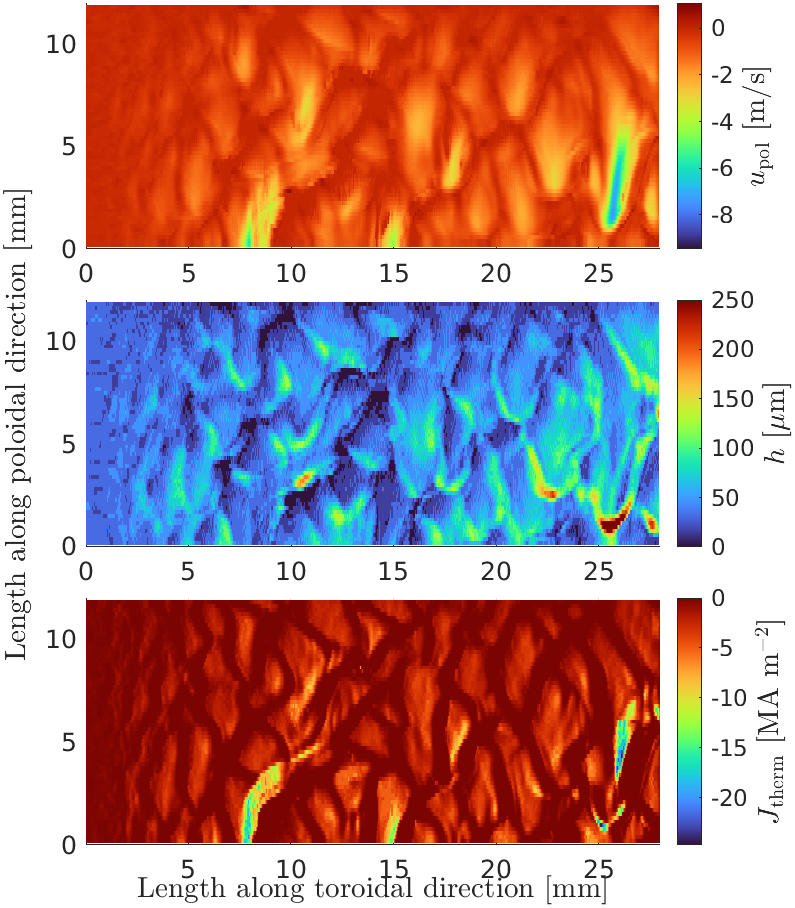}
    \caption{Results of 3D runs (with melt motion as well as with coupled heat flux and deformation) for the worse case loading scenario, $f_\mathrm{ELM}=10\,$Hz and $q_\parallel^{\mathrm{inter}} = 6/\sin(a_0)$\,[MWm$^{-2}$]. The poloidal velocity profile ($u_\mathrm{pol}$), melt depth ($h$), and escaping current density ($J_\mathrm{esc}$) $1\,$ms after the ELM at $5\,$s impacts the MB.}
    \label{fig:ELM_quantites}
\end{figure}

Aiming to facilitate the discussion of the macroscopic melt motion, the melt depth, the melt velocity and the escaping thermionic current density are illustrated in Fig.\ref{fig:ELM_quantites} for the instance of 1\,ms after the ELM at 5\,s impacted the MB. The simulations reveal that for most melt pools and during most of their lifetime, the thermionic emission is limited and does not follow the unimpeded Richardson-Dushman expression. Consequently, the escaping current depends on the magnetic field inclination angle and not on the surface temperature. The profiles of all these quantities are elaborate, featuring fine spatial variations - striations.  The melt dynamics vary spatially due to \textbf{(i)} variations of the melt depth $h$ that result in changes of the viscous damping that scale as $\propto{h}^{-2}$, \textbf{(ii)} variations of the escaping current densities which scale as $\propto\sin^2{a}$, and give rise to a non-uniform driving Lorenz force, \textbf{(iii)} the presence of shadowed regions, where the pool lifetime is shorter due to the faster inter-ELM resolidification. These factors operate in a synergistic fashion. Namely, in regions where $a$ is closer to $\pi/2$, i.e. non-shadowed regions where the heat flux is incident on the top surface with a large angle, the pools are deeper and hence less damped, while the forcing is stronger and acts for a longer time resulting in the observed elongated deformation patterns. 

\begin{figure}
    \centering
    \includegraphics[width=1.0\linewidth]{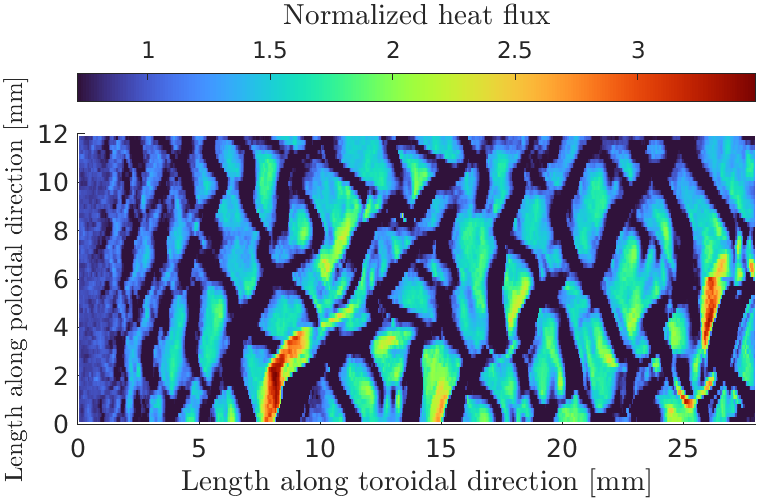}
    \caption{Results of 3D runs (with melt motion as well as with coupled heat flux and deformation) for the worse case loading scenario, $f_\mathrm{ELM}=10\,$Hz and $q_\parallel^{\mathrm{inter}} = 6/\sin(a_0)$\,[MWm$^{-2}$]. The normalized heat flux deposited on the deformed sample after $5\,$s of exposure. Normalization to the heat loading for the pristine sample, i.e. for decoupled deformation and loading by setting $F_{\mathrm{o}} = 0$ in Eq.\eqref{eqn:par2perp}.}
    \label{fig:normalized_flux}
\end{figure}

Essentially, the striations and non-uniform melt dynamics stem from the coupling between the heat flux projection and the surface deformation. Fig.\ref{fig:normalized_flux} shows the spatial variations of the loading by normalizing to the heat flux that would have been deposited on the initial pristine surface, i.e., setting $F_{\mathrm{o}}=0$ in Eq.\eqref{eqn:par2perp}. The normalized projected heat flux reveals the effective spatial variations of the heat flux loading. The integrated value of the peak power per ELM in Fig.\ref{fig:energy_balance} changes only up to $\sim15\%$ during the simulation. Meanwhile, the heat flux at the instant shown in Fig.\ref{fig:normalized_flux} varies by a factor of more than 5 at different locations and up to $\sim3.5$ when compared to the projection on the pristine surface. These spatial variations are driving the overall dynamics and ultimately result in the striations formed in the final deformation profile.

\subsubsection{Constant loading.}\label{sssec:constant_loading}

Aiming to further highlight the effect of coupling between the heat flux projection and surface deformation, the above scenario is also simulated under constant loading, where the heat flux projection is given by the inclination angle with respect to the pristine surface $a_0$ (i.e., setting $F_{\mathrm{o}} = 0$ in Eq.\eqref{eqn:par2perp}). In such loading case, after $5\,$s, the cavity and hill volumes reach 1.5\,mm$^3$ and 0.4\,mm$^3$, respectively. These are to be compared to the respective values in the reference case, 32\,mm$^3$ and 7\,mm$^3$. Note that the cavity and hill volumes are not equal. At each ELM, the full top surface melts and the Lorentz force drives material outside the computational domain as a result of the free outflow boundary condition implemented in MEMENTO.

There is a drastic increase in the melt volume when the heat flux projection takes surface deformation into account, despite the fact that the energy deposited over the first $5\,$s was merely $\sim6\%$ higher in the reference case compared to the constant loading case (in which the projection angles due to deformation are not accounted for in deriving the incident heat flux). Previous W melting investigations showed that minor changes in the heat flux can cause substantial changes in the generated melt\,\cite{Thoren2018,  Ratynskaia2020, Thoren2021, Ratynskaia2021, Ratynskaia2022}. This is due to the fact that most energy is expended on heating up the bulk and only a small fraction is expended on the enthalpy of fusion. Here, the heat flux magnitude can vary by more than a factor of 2 between the two cases, as discerned by the time-instant presented in Fig.\ref{fig:normalized_flux}. Such spatial variation of the heat flux is crucial for the formation of melt and the final deformation profile, especially since the variations occur over distances larger than the heat diffusion length during the pool lifetime ($\sim0.12\,$mm for $\sim1$\,ms.)

Despite the realization of full top surface melting at each ELM in both cases, the melt dynamics is very different depending on whether the coupling of heat flux and surface deformation is taken into account or not. With constant heat flux, the pool depth and lifetime variations at a given time are dictated exclusively by the distance of the solid-liquid interface from the cooling pipe. For example, the ELM striking at the $5\,$s instant creates $\sim$30 $\mu$m deep pools at the center of the top surface, but $\sim$40 $\mu$m and $\sim$50 $\mu$m deep pools at the edges. The edge asymmetry is a consequence of bevelling. The melt dynamics in case of $F_{\mathrm{o}} = 2/7$ is considerably more complex, as discussed in Sec.\ref{sssec:coupling}.

\subsubsection{Uncertainty in thermionic emission.}\label{sssec:thermionic}

In order to investigate the impact of escaping current uncertainties (introduced by the effect of the non-planarity of the exposed surface and the unknown plasma conditions in front of the shadowed regions), two limiting cases are compared. In the reference case, the escaping current density in the shadowed regions was set to zero, while, in the test case, the expression valid for non-shadowed regions will be employed all over the top surface. 

In terms of the deformed volume, the two cases differ by merely $\sim15\%$. When the escaping current density is set to zero at the shadowed free surface regions, two competing effects locally modify the melt dynamics. The shadowed side of the pools is no longer cooled by thermionic emission, thus the lifetime is increased. On the other hand, there is also a smaller replacement current through the melt pool, and hence a weaker Lorentz force driving the displacement. This cancellation of effects, along with the fact that the shadowed regions do not significantly contribute to the displaced volume due to their shorter lifetime, results in the small overall difference between the two cases.

\subsubsection{Effect of increased inter-ELM heat fluxes.}\label{sssec:interELM}

\begin{figure}
    \addtolength{\tabcolsep}{-6pt} 
    \begin{tabular}{c}
        \subfloat{%
          \begin{overpic}[width = 3.3in]{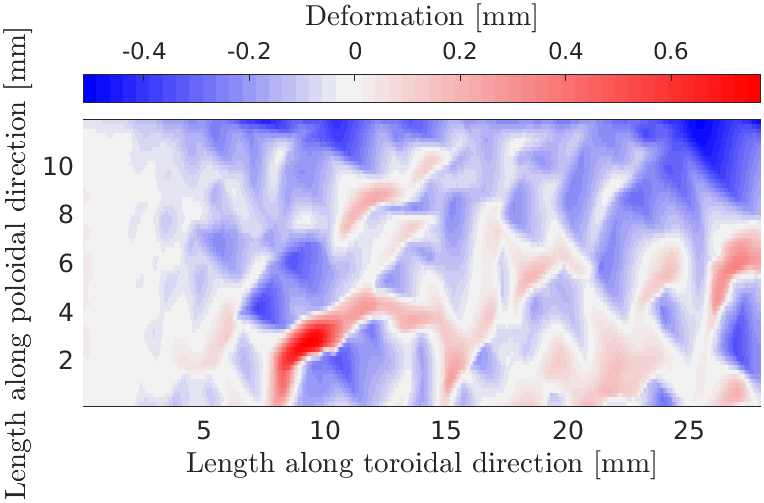}
          \end{overpic}
        }\\
        \subfloat{%
          \begin{overpic}[width = 3.3in]{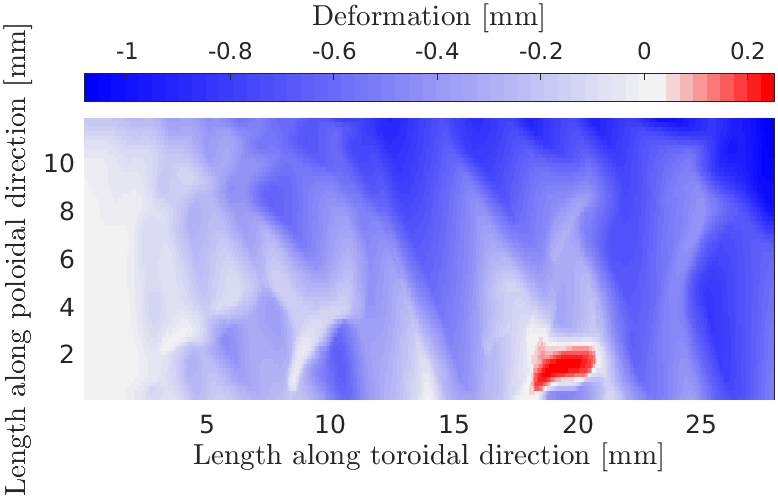}
          \end{overpic}
        }
    \end{tabular}
    \addtolength{\tabcolsep}{6pt} 
    \caption{Results of 3D runs (with melt motion as well as with coupled heat flux and deformation) for the worse case loading scenario, $f_\mathrm{ELM}=10\,$Hz, $q_\parallel^{\mathrm{inter}} = 6/\sin(a_0)$\,[MWm$^{-2}$] (top) and $q_\parallel^{\mathrm{inter}} = 12/\sin(a_0)$\,[MWm$^{-2}$] (bottom). The surface deformation profile after 5\,s of exposure. Note the different color bar in the upper and lower panels.}\label{fig:inter_elm}
\end{figure}

The increase of the inter-ELM heat flux magnitude strongly affects the deformation volume in the first few seconds compared to the reference case due to different melt depths and lifetimes. After $5\,$s of loading with a $2\times$ higher inter-ELM of $12/\sin\,a_0$\,[MW/m$^2$], the volume of the craters reaches 142 mm$^3$. This is to be compared to 32\,mm$^3$ in the reference case. Such a difference is clearly observed in the deformation profiles depicted in Fig.\ref{fig:inter_elm}.

The inter-ELM heat flux increase does not suffice for sustained melting to be realized, but the pool lifetime increases, because the mechanisms that promote re-solidification, heat conduction and surface cooling are suppressed. For higher inter-ELM heat fluxes, the elevated temperatures in the bulk near the free surface result in weaker conductive cooling for the pools. Indeed, the surface temperatures before the ELM impacts are $\sim2000-2400$\,K in the reference case and $\sim2100-2800$\,K for the augmented inter-ELM heat flux. In addition, higher inter-ELM heat fluxes can more readily compensate for the effect of surface cooling. 

\begin{figure}
    \includegraphics[width=1.0\linewidth]{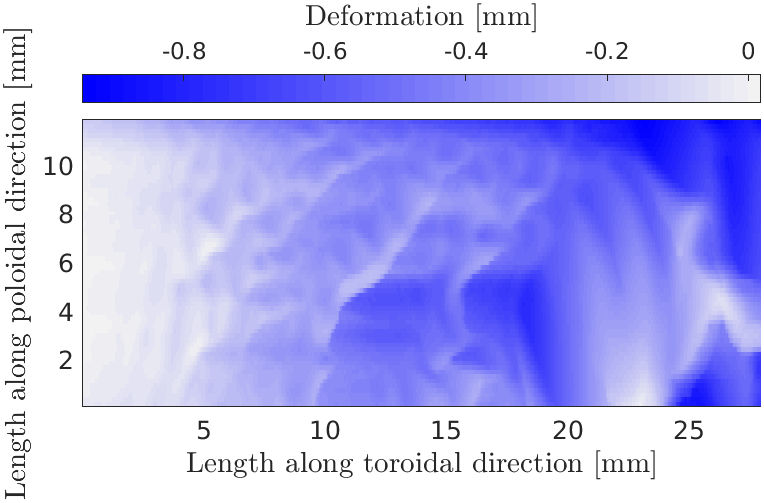}
    \caption{Results of 3D runs (with melt motion as well as with coupled heat flux and deformation) for the worse case loading scenario, $f_\mathrm{ELM}=10\,$Hz, $q_\parallel^{\mathrm{inter}} = 6/\sin(a_0)$\,[MWm$^{-2}$]. The surface deformation profile after $10\,$s of exposure. Note the different color bar compared to Fig.\ref{fig:inter_elm}.}\label{fig:ref_10s}
\end{figure}

It should be reiterated that after the top surface is distorted, the deformation volume quickly builds up even in the lower inter-ELM run, see Sec.\ref{sssec:coupling}. Hence, the melt damage caused by the higher inter-ELM heat fluxes after $5\,$s will also be reached in the lower case, just at a later instant. To be quantitative, after $10\,$s of loading with the reference inter-ELM heat flux, the crater volumes grow to 120 mm$^3$ and the deformation profiles become similar, compare Fig.\ref{fig:inter_elm} (lower) with Fig.\ref{fig:ref_10s}. 

A similar conclusion is also valid for loading with smaller ELMs, e.g. those that in the 2D runs resulted in no melting. Had such ELMs impacted an already deformed surface, such as the one presented in Fig.\ref{fig:inter_elm}, and scaled as suggested by Fig.\ref{fig:normalized_flux}, then those ELMs could aggravate the already damaged surface. Overall, if sufficiently strong ELMs alter the top surface, then the MB is susceptible to melting even from ELMs that would not otherwise melt the pristine surface.


\section{Summary and conclusions}\label{sec:summary}

Modelling of the computationally challenging scenario of macroscopic melt motion induced on actively cooled ITER W monoblocks by ELM heat loads has been successfully performed with the new MEMENTO melt dynamics code. The simulations were rendered feasible courtesy of very recent numerical updates\,\cite{Paschalidis2023,Paschalidis2024}.

\begin{table}[]
\begin{center}
\begin{tabular}{c|ccccc}
\hline
Case                        & $f_\mathrm{ELM}$ & $T_\mathrm{s}$ & $T_\mathrm{2mm}$ & $T_\mathrm{pipe}$ & h \\ \hline
\multirow{2}{*}{Nominal} & 1\,Hz             & 3732\,K         & 957\,K            & 508\,K             & 8\,$\mu$m  \\
                         & 10\,Hz            & 1944\,K         & 1065\,K           & 541\,K             & 0\,$\mu$m \\ \hline
\multirow{5}{*}{Worst}   & 5\,Hz             & 5714\,K         & 1602\,K           & 560\,K             & 65\,$\mu$m \\
                         & 10\,Hz            & 4563\,K         & 1772\,K           & 565\,K             & 33\,$\mu$m \\
                         & 15\,Hz            & 4043\,K         & 1877\,K           & 572\,K             & 13\,$\mu$m \\
                         & 20\,Hz            & 3871\,K         & 1967\,K           & 575\,K             & 6\,$\mu$m \\
                         & 50\,Hz            & 3695\,K         & 2348\,K           & 603\,K             & $<6\,\mu$m \\ \hline
\end{tabular}
\caption{Summary of the 2D simulations without melt motion presented in Sec.\ref{ssec:2d_runs}. The tabulated temperatures correspond to the maximum values reached at the middle of the loaded surface ($T_\mathrm{s}$), at $2\,$mm below ($T_\mathrm{2mm}$) and at the coolant interface point closest to the free surface ($T_\mathrm{pipe}$). The maximum melt depth ($h$) for each case is also provided. In the worst case with $f_\mathrm{ELM} = 50\,$Hz, the melt depth is smaller than the discretization length. The nominal and worst case heat fluxes are discussed in Sec.\ref{ssec:implem_loading}. CHF events are assumed to occur for $T_\mathrm{pipe}>573\,$K.}
\label{tab:2D_results}
\end{center}
\end{table}

The ELM parameter space was investigated to identify operational windows in which MB damage occurs due to melt motion, recrystallization and CHF events. A summary is presented in Table \ref{tab:2D_results}. In particular, it is predicted that there is marginal melting with $q^\mathrm{inter}_{||} = 6/\sin{a_0} \,$MW/m$^2$, $f_{\mathrm{ELM}}=1\,$Hz and $\mathrm{max}\left( q_{||} \right)\approx 65$\,GW/m$^2$, while there is no melting for $q^\mathrm{inter}_{||} = 6/\sin{a_0}\,$MW/m$^2$, $f_{\mathrm{ELM}}=10\,$Hz and $\mathrm{max}\left( q_{||} \right) \approx 20$\,GW/m$^2$. At the upper threshold of the uncertainties of the Eich scaling, melting was observed for all ELM frequencies investigated. The evolution of the crater volumes is summarized in Table \ref{tab:3D_results}. Notably, for $f_{\mathrm{ELM}}=10\,$Hz, $q^\mathrm{inter}_{||} = 6/\sin{a_0}\,$MW/m$^2$, and $\mathrm{max}\left( q_{||} \right)\approx 65$\,GW/m$^2$, appreciable damage with 2.3\,mm$^3$ crater volume is reached after $2\,$s, which increases to $32\,$mm$^3$ by $5\,$s. Doubling the perpendicular inter-ELM flux, keeping all other parameters constant, accelerates the deformation build-up, with $142\,$mm$^3$ for the crater volume by $5\,$s. With smaller inter-ELM values significant damage is not entirely avoided, but it is merely postponed to later instants.

\begin{table}[]
\begin{center}
\begin{tabular}{l|llll}
\hline
Scenario            & 2\,mm$^3$      & 10\,mm$^3$      & 50\,mm$^3$                \\ \hline
Reference           & 1.9\,s         & 3.5\,s          & 5.7\,s                    \\
Const. loading      & 5.7\,s         & $>$10\,s        & $>$10\,s                  \\
Emitting shadows    & 1.9\,s         & 3.3\,s          & 5.4\,s                    \\
Double inter-ELM    & 1.4\,s         & 2.2\,s          & 3.4\,s                    \\ \hline
\end{tabular}
\caption{Summary of the evolution of the crater volume from the 3D simulations including melt motion. The reference case is analysed in Sec.\ref{sssec:coupling}, the constant loading case in Sec.\ref{sssec:constant_loading}, the case with emitting shadows in Sec.\ref{sssec:thermionic} and the double inter-ELM heat flux in Sec.\ref{sssec:interELM}. Note that, for constant loading, the crater volume does not reach 10\,mm$^3$ during the simulated 10\,s.}
\label{tab:3D_results}
\end{center}
\end{table}

The importance of the coupling between heat flux loading and surface deformation was highlighted. The interplay is present for all shallow incidence angle cases. It stems from the decrease in the projection area for the parallel flux due to surface deformation and it results in a positive feedback loop where already deformed surfaces are subject to even stronger heat fluxes. This coupling implies that safe ELMy operational windows are history-dependent. ITER ELMs that cannot melt a pristine MB could aggravate the damage on an already deformed MB. Consequently, even small ELMs can amount to significant deformation, if a series of strong ELMs have initially modified the pristine top surface. 

For each scenario, a single set of ELM parameters was used to generate a fully periodic parallel heat flux that is wetting the MB top surface. Attempting to statistically represent a realistic sequence of small and large ELMs would create a unique deformation profile. However, given that the ELM statistical behaviour is largely unknown even in existing tokamaks, the construction of a particular ELM trace would not bring any further insight to the problem. Neither would introducing such complexities change the key message of the current work;  loading with smaller ELMs, even the ones considered safe for the pristine surface, slows down the damage progression but does not fully prevent it.

\section*{Acknowledgments}

\noindent SR and PT acknowledge the financial support of the Swedish Research Council under Grant No\,2021-05649. This work has received funding from the ITER Organization within the auspices of the Implementing Agreement No.\,1 (Ref.IO/IA/22/4300002750) to the Agreement on Academic and Scientific Cooperation (Ref.LGA-2021-A-75) between ITER and KTH. ITER is the Nuclear Facility INB No.\,174. This publication is provided for scientific purposes only and its contents should not be considered as commitments from the ITER Organization as a nuclear operator in the frame of the licensing process. The views and the opinions expressed herein do not necessarily reflect those of the ITER Organization. The MEMENTO simulations were partly enabled by resources provided by the National Academic Infrastructure for Supercomputing in Sweden (NAISS) at the NSC (Link\"oping University) partially funded by the Swedish Research Council through grant agreement No\,2022-06725.\\

\bibliography{biblio_new}

\begin{thebibliography}{10}

\bibitem{Pitts2017}
R.~A. Pitts, S.~Bardin, B.~Bazylev, M.~A. {van den Berg}, et~al.
\newblock Physics conclusions in support of {{ITER W}} divertor monoblock shaping.
\newblock {\em Nucl. Mater. Energy}, 12:60, 2017.

\bibitem{Ratynskaia2020}
S.~Ratynskaia, E.~Thor{\'e}n, P.~Tolias, R.~A. Pitts, et~al.
\newblock Resolidification-controlled melt dynamics under fast transient tokamak plasma loads.
\newblock {\em Nucl. Fusion}, 60:104001, 2020.

\bibitem{Tsuchida2018}
K.~Tsuchida, T.~Miyazawa, A.~Hasegawa, S.~Nogami, and M.~Fukuda.
\newblock Recrystallization behavior of hot-rolled pure tungsten and its alloy plates during high-temperature annealing.
\newblock {\em Nucl. Mater. Energy}, 15:158, 2018.

\bibitem{Escourbiac2019}
F.~Escourbiac, A.~Durocher, A.~Fedosov, T.~Hirai, et~al.
\newblock Assessment of critical heat flux margins on tungsten monoblocks of the {ITER} divertor vertical targets.
\newblock {\em Fus. Eng. Des.}, 146:2036, 2019.

\bibitem{Hirai2016}
T.~Hirai, S.~Panayotis, V.~Barabash, C.~Amzallag, et~al.
\newblock Use of tungsten material for the {ITER} divertor.
\newblock {\em Nucl. Mater. Energy}, 9:616, 2016.

\bibitem{Hirai2018}
T.~Hirai, S.~Carpentier-Chouchana, F.~Escourbiac, S.~Panayotis, et~al.
\newblock Design optimization of the {ITER} tungsten divertor vertical targets.
\newblock {\em Fus. Eng. Des.}, 127:66, 2018.

\bibitem{Panayotis2017}
S.~Panayotis, T.~Hirai, V.~Barabash, C.~Amzallag, et~al.
\newblock Fracture modes of {ITER} tungsten divertor monoblock under stationary thermal loads.
\newblock {\em Fus. Eng. Des.}, 125:256, 2017.

\bibitem{Gunn2017}
J.~P. Gunn, S.~Carpentier-Chouchana, F.~Escourbiac, T.~Hirai, et~al.
\newblock Surface heat loads on the {ITER} divertor vertical targets.
\newblock {\em Nucl. Fusion}, 57:046025, 2017.

\bibitem{Van_den_Kerkhof2021}
S.~{Van den Kerkhof}, M.~Blommaert, R.~A. Pitts, W.~Dekeyser, S.~Carli, and M.~Baelmans.
\newblock Impact of {ELM} mitigation on the {ITER} monoblock thermal behavior and the tungsten recrystallization depth.
\newblock {\em Nucl. Mater. Energy}, 27:101009, 2021.

\bibitem{Krieger2018}
K.~Krieger, M.~Balden, J.~W. Coenen, F.~Laggner, et~al.
\newblock Experiments on transient melting of tungsten by {{ELMs}} in {{ASDEX Upgrade}}.
\newblock {\em Nucl. Fusion}, 58:026024, 2018.

\bibitem{Thoren2018}
E.~Thor{\'e}n, S.~Ratynskaia, P.~Tolias, R.~A. Pitts, K.~Krieger, M.~Komm, and M.~Balden.
\newblock {MEMOS 3D} modelling of {ELM}-induced transient melt damage on an inclined tungsten surface in the {ASDEX Upgrade} outer divertor.
\newblock {\em Nucl. Mater. Energy}, 17:194, 2018.

\bibitem{Paschalidis2023}
K.~Paschalidis, S.~Ratynskaia, F.~Lucco~Castello, and P.~Tolias.
\newblock Melt dynamics with {MEMENTO} -- code development and numerical benchmarks.
\newblock {\em Nucl. Mater. Energy}, 37:101545, 2023.

\bibitem{Paschalidis2024}
K.~Paschalidis, F.~Lucco~Castello, S.~Ratynskaia, P.~Tolias, and L.~Brandt.
\newblock The {MEMENTO} code for modelling of macroscopic melt motion in fusion devices.
\newblock {\em Fus. Eng. Des. [submitted]}, 2024.

\bibitem{Thoren2017}
E.~Thor{\'e}n, B.~Bazylev, S.~Ratynskaia, et~al.
\newblock Simulations with current constraints of {ELM}-induced tungsten melt motion in {ASDEX Upgrade}.
\newblock {\em Phys. Scr.}, T170:014006, 2017.

\bibitem{Thoren2018_1}
E.~Thor{\'e}n, P.~Tolias, S.~Ratynskaia, R.~A. Pitts, and K.~Krieger.
\newblock Self-consistent description of the replacement current driving melt layer motion in fusion devices.
\newblock {\em Nucl. Fusion}, 58:106003, 2018.

\bibitem{Thoren2021}
E.~Thor{\'e}n, S.~Ratynskaia, P.~Tolias, and R.~A. Pitts.
\newblock The {MEMOS-U} code description of macroscopic melt dynamics in fusion devices.
\newblock {\em Plasma Phys. Control. Fusion}, 63:035021, 2021.

\bibitem{Ratynskaia2021}
S.~Ratynskaia, E.~Thor{\'e}n, P.~Tolias, R.~A. Pitts, and K.~Krieger.
\newblock The {MEMOS-U} macroscopic melt dynamics code — benchmarking and applications.
\newblock {\em Phys. Scr.}, 96:124009, 2021.

\bibitem{Ratynskaia2022}
S.~Ratynskaia, K.~Paschalidis, P.~Tolias, K.~Krieger, et~al.
\newblock Experiments and modelling on {ASDEX} upgrade and {WEST} in support of tool development for tokamak reactor armour melting assessments.
\newblock {\em Nucl. Mater. Energy}, 33:101303, 2022.

\bibitem{Ratynskaia2024}
S.~Ratynskaia, K.~Paschalidis, K.~Krieger, L.~Vignitchouk, et~al.
\newblock Metallic melt transport across castellated tiles.
\newblock {\em Nucl. Fusion}, 64:036012, 2024.

\bibitem{Smid1998}
I.~Smid, M.~Akiba, G.~Vieider, and L~Plöchl.
\newblock Development of tungsten armor and bonding to copper for plasma-interactive components.
\newblock {\em J. Nucl. Mater.}, 258-263:160, 1998.

\bibitem{Davis1998}
J.~W. Davis, V.~R. Barabash, A.~Makhankov, L.~D. Plochl, and K.~T. Slattery.
\newblock Assessment of tungsten for use in the {ITER} plasma facing components.
\newblock {\em J. Nucl. Mater.}, 258-263:160, 2017.

\bibitem{Li2015}
M.~Li and J.~H. You.
\newblock Interpretation of the deep cracking phenomenon of tungsten monoblock targets observed in high-heat-flux fatigue tests at {20 MW/m2}.
\newblock {\em Fus. Eng. Des.}, 101:1, 2015.

\bibitem{Li2017}
M.~Li and J.~H. You.
\newblock Design options to mitigate deep cracking of tungsten armor.
\newblock {\em Fus. Eng. Des.}, 124:468, 2017.

\bibitem{Pitts2019}
R.~A. Pitts, X.~Bonnin, F.~Escourbiac, H.~Frerichs, J.~P. Gunn, et~al.
\newblock Physics basis for the first {{ITER}} tungsten divertor.
\newblock {\em Nucl. Mater. Energy}, 20:100696, 2019.

\bibitem{Raffray1999}
A.R. Raffray, J.~Schlosser, M.~Akiba, M.~Araki, S.~Chiocchio, D.~Driemeyer, F.~Escourbiac, S.~Grigoriev, M.~Merola, R.~Tivey, G.~Vieider, and D.~Youchison.
\newblock Critical heat flux analysis and {R\&D} for the design of the {ITER} divertor.
\newblock {\em Fus. Eng. Des.}, 45:377, 1999.

\bibitem{Boscary1999}
J.~Boscary, J.~Fabre, and J.~Schlosser.
\newblock Critical heat flux of water subcooled flow in one-side heated swirl tubes.
\newblock {\em Int. J. Heat Mass Transf.}, 42:287, 1999.

\bibitem{Loarte2003}
A.~Loarte, G.~Saibene, R.~Sartori, M.~Becoulet, et~al.
\newblock {ELM} energy and particle losses and their extrapolation to burning plasma experiments.
\newblock {\em J. Nucl. Mater.}, 313-316:962, 2003.

\bibitem{Loarte2007}
A.~Loarte, G.~Saibene, R.~Sartori, V.~Riccardo, et~al.
\newblock Transient heat loads in current fusion experiments, extrapolation to {ITER} and consequences for its operation.
\newblock {\em Phys. Scr.}, T128:222, 2007.

\bibitem{Loarte2014}
A.~Loarte, G.~Huijsmans, S.~Futatani, L.~R. Baylor, et~al.
\newblock Progress on the application of {ELM} control schemes to {ITER} scenarios from the non-active phase to {DT} operation.
\newblock {\em Nucl. Fusion}, 54:033007, 2014.

\bibitem{Fundamenski2006}
W.~Fundamenski and R.~A. Pitts.
\newblock A model of {ELM} filament energy evolution due to parallel losses.
\newblock {\em Plasma Phys. Control. Fusion}, 48:109, 2006.

\bibitem{Eich2017}
T.~Eich, B.~Sieglin, A.~J. Thornton, M.~Faitsch, et~al.
\newblock {ELM} divertor peak energy fluence scaling to {ITER} with data from {JET}, {MAST} and {ASDEX} upgrade.
\newblock {\em Nucl. Mater. Energy}, 12:84, 2017.

\bibitem{Komm2017}
M.~Komm, S.~Ratynskaia, P.~Tolias, et~al.
\newblock On thermionic emission from plasma-facing components in tokamak-relevant conditions.
\newblock {\em Plasma Phys. Control. Fusion}, 59:094002, 2017.

\bibitem{Komm2017_1}
M.~Komm, P.~Tolias, S.~Ratynskaia, et~al.
\newblock Simulations of thermionic suppression during tungsten transient melting experiments.
\newblock {\em Phys. Scr.}, T 170:014069, 2017.

\bibitem{Komm2020}
M.~Komm, S.~Ratynskaia, P.~Tolias, and A.~Podolnik.
\newblock Space-charge limited thermionic sheaths in magnetized fusion plasmas.
\newblock {\em Nucl. Fusion}, 60:054002, 2020.

\bibitem{Tolias2020}
P.~Tolias, M.~Komm, S.~Ratynskaia, and A.~Podolnik.
\newblock Origin and nature of the emissive sheath surrounding hot tungsten tokamak surfaces.
\newblock {\em Nucl. Mater. Energy}, 25:100818, 2020.

\bibitem{Tolias2023}
P.~Tolias, M.~Komm, S.~Ratynskaia, and A.~Podolnik.
\newblock {ITER} relevant multi-emissive sheaths at normal magnetic field inclination.
\newblock {\em Nucl. Fusion}, 63:026007, 2023.

\bibitem{Tolias2014_1}
P.~Tolias.
\newblock On secondary electron emission and its semi-empirical description.
\newblock {\em Plasma Phys. Control. Fusion}, 56:123002, 2014.

\bibitem{Tolias2014}
P.~Tolias.
\newblock On electron backscattering from dust grains in fusion plasmas.
\newblock {\em Plasma Phys. Control. Fusion}, 56:045003, 2014.

\bibitem{Skovorodin2016}
D.~I. Skovorodin, A.~A. Pshenov, A.~S. Arakcheev, E.~A. Eksaeva, E.~D. Marenkov, and S.~I. Krasheninnikov.
\newblock Vapor shielding models and the energy absorbed by divertor targets during transient events.
\newblock {\em Phys. Plasmas}, 23:022501, 2016.

\end{thebibliography}
\end{document}